\documentclass[preprint,10pt]{llncs}
\usepackage{color,graphicx}
\usepackage{amsmath}
\usepackage{amssymb}
\usepackage{bm}                 
\usepackage{mathrsfs}           
\usepackage{dsfont}             
\usepackage{mathtools}

\newcommand\NN{{\mathds{N}}}
 
\newcommand\RR{{\mathds{R}}}

\newcommand{\avg}[1]{\left\langle #1 \right\rangle}
\newcommand{\norm}[1]{\left\|#1\right\|}

\newcommand\rank{\mathop{rank}}
\newcommand\codim{\mathop{codim}}
\newcommand\eps{{\varepsilon}}
\newcommand\tsp{\intercal}      
\newcommand\rmd{{\mathrm d}}    

\newcommand\St{{\sf S}}           
\newcommand\vX{{{\bf X}}}
\newcommand\vx{{{\bf x}}}
\newcommand\vY{{{\bf Y}}}
\newcommand\vy{{{\bf y}}}
\newcommand\vZ{{{\bf Z}}}
\newcommand\vz{{{\bf z}}}

\newcommand\vnu{{\bm\nu}}
\newcommand\vmu{{\bm\mu}}
\newcommand\vsg{{\bm\sigma}}
\newcommand{\PP}[1]{\mathds{P}\left\{#1\right\}}
\newcommand{\EE}{\mathds{E}}
\newcommand{\calX}{\mathcal{X}}
\newcommand{\sref}[1]{section \ref{#1}}

\newcounter{asmp}
\newenvironment{assumption}
{\par\medskip\noindent\refstepcounter{asmp}\hbox{\bf Assumption \arabic{asmp}.} }
{\par\medskip}

\title{Mean-Field approximation and Quasi-Equilibrium reduction of Markov Population Models\thanks{We acknowledge partial support from EU-FET project QUANTICOL (nr. 600708) and  FRA-UniTS.}}
\author{Luca Bortolussi\inst{1,2} \and Rytis Pa\v{s}kauskas\inst2}
\institute{Department of Mathematics and  Geosciences, University of Trieste \and 
  ISTI Area della Ricerca CNR, via G. Moruzzi 1, 56124 Pisa, Italy}

\begin{document}
\maketitle

\begin{abstract}
  Markov Population Model is a commonly used framework to describe stochastic systems. Their exact analysis is unfeasible in most cases because of the state space explosion. Approximations are usually sought, often with the goal of reducing the number of variables. Among them, the mean field limit and the quasi-equilibrium approximations stand out. We view them as techniques that are rooted in independent basic principles. At the basis of the mean field limit is the law of large numbers. The principle of the quasi-equilibrium reduction is the separation of temporal scales. It is common practice to apply both limits to an MPM yielding a fully reduced model. Although the two limits should be viewed as completely independent options, they are applied almost invariably in a fixed sequence: MF limit first, QE-reduction second. We present a framework that makes explicit the distinction of the two reductions, and allows an arbitrary order of their application. By inverting the sequence, we show that the double limit does not commute in general: the mean field limit of a time-scale reduced model is not the same as the time-scale reduced limit of a mean field model. An example is provided to demonstrate this phenomenon. Sufficient conditions for the two operations to be freely exchangeable are also provided. 
\end{abstract}

\date{\today}

\section{Introduction}\label{s:intro}
Many complex systems whose dynamics is the result of the interaction of populations of indistinguishable agents can be described by Markov Population Models (MPM, \cite{Bortolussi2013,HenzingerWolf2011}). This is the case, for instance, for biological systems and computer systems like queuing networks. Quantitative formal methods offer a powerful framework to describe and analyse them, using tools from verification and model checking. However, formal analysis of the Continuous Time Markov Chain (CTMC) that underlies an MPM is extremely challenging due to its usually large state space. Approximation techniques are therefore extremely useful, as they can lead to considerable simplifications of the analysis phase. 

In this paper, we discuss two such methods. The first one is the fluid or mean-field approximation \cite{Bortolussi2013,EthierKurtz2005,BenaimLeBoudec2008}. It has received considerable attention in the quantitative formal methods community in the past years with applications also to passage time computations \cite{hayden_performance_2013} and stochastic model checking \cite{bortolussi_fluid_2012,bortolussiLanciani_2013}. This method is based on a version of the law of large numbers for stochastic processes, known as Kurtz' theorem \cite{EthierKurtz2005} which guarantees that, for large populations, an MPM is close to a (deterministic) ordinary differential equation (called fluid or mean-field ODE), converging to the latter in the limit of infinite population. This approximation holds for the transient behaviour under mild conditions on rate functions and model transitions, and can be extended to the steady state behaviour under additional assumptions on the limit ODEs \cite{benaim_deterministic_2003,benaim_mean_2011}. 

Multiple time scale reduction, on the other hand, is based on a common intrinsic property of multi-dimensional dynamical systems to equilibrate unevenly. Several dimensions can be removed from a model if certain degrees of freedom equilibrate much faster than the rest. This is achieved by identifying the fast components and approximating them with the conditional equilibrium distribution. A rigorous definition involves singular perturbation \cite{Verhulst2005}: A model that is a singular perturbation of another model is a multi-scale model. But this definition is too restrictive for real life situations (models with numerical rate constants of the same order may be multi-scale). Several methodologies and criteria to detect multiple time scales have been developed over the years: quasi-equilibrium and quasi-stationary state \cite{reduction-mf}, computational singular perturbation\cite{lam}, intrinsic low dimensional manifold \cite{maas} etc. Most of them have originated in chemistry \cite{reduction-mf,maas,lam} (ODE models) and \cite{reduction-me,mastny_two_2007,cao,weinan_nested_2007} (stochastic models) and quite often the impression is that ODE and stochastic reductions are based on different assumptions. In the present article we propose a framework that eliminates this prejudice for a special, but important class of so-called quasi-equilibrium reductions \cite{BortPaskSurvey14}.
Our contributions can be summarised as follows:
\begin{itemize}
\item we provide a consistent and constructive definition of the quasi-equilibrium reduction for MPMs. In particular, we treat uniformly mean field equations and stochastic processes by constructing reductions at a level of the MPM formalism \cite{HenzingerWolf2011}. For the stochastic case, we also formally prove the convergence of the full model to the reduced one when fast and slow time scales diverge.
\item by examining the relationship between the QE reduction of the MF limit of a population process and the MF limit of the QE-reduced stochastic system we give sufficient conditions for the mean-field limit of the reduced stochastic system to exist and to be equal to the reduced mean field model. We also show that this is not true in general, and discuss scenarios where, application of the two limits in different order results in non-equivalent approximations. 
\end{itemize}

The paper is organised as follows: In Section \ref{s:mpm}, we introduce Markov Population Models, while in Section \ref{s:mf} we review the mean-field approximation. Section \ref{s:qe}, instead, is devoted to the presentation of the Quasi-Equilibrium reduction, both for differential equations and for MPMs. Section \ref{s:comp} contains our results about the relationship between mean-field and quasi-equilibrium, while in Section \ref{s:dis} we draw the final conclusions.  

\section{Markov Population Models}\label{s:mpm}

A  Markov Population Model \cite{HenzingerWolf2011,Bortolussi2013} is a simple formalism to describe  models of populations of interacting agents based on  Continuous-Time Markov Chains (CTMC). The formalism is inspired by chemical reaction networks \cite{Gillespie1977JPC}, and is formally characterised by a tuple $\calX=(\vX, {\mathcal M}, {\mathcal T}, \vX_0)$ where
\begin{enumerate}
\item $\vX=(X_1,\dots, X_n)^\tsp$ is a (column) vector of variables describing the $n$ species of the model.
\item ${\mathcal M}$ is the domain of $\vX$. Usually $X_i$ counts the number of elements in a population of a species, therefore we assume $X_i\in\NN$ and ${\mathcal M}\subseteq\NN^n$.
\item ${\mathcal T}=\{\tau_1,\dots,\tau_r\}$ is the set of $r$ \emph{transitions}, of the form $\tau=(\vnu,W)$, where:
  \begin{enumerate}
  \item $\vnu=(\nu_1,\dots,\nu_n)^\tsp \in{\mathcal M}$ is a (column) \emph{update vector}. This vector determines the stoichiometry of a transition, i.e. its elements equal the net change of the corresponding variable due to the transition.
  \item $W:{\mathcal M}\mapsto{\RR}_{\geq 0}$ is the \emph{rate function}. We impose that all rate functions satisfy  $W(\vX)\ge 0$ and $W(\vX)=0$ if $\vX+\vnu\notin{\mathcal M}$.
 \end{enumerate}
\item $\vX_0\in{\mathcal M}$ is the initial state:  the process starts in $\vX_0$ with probability one.
\end{enumerate}

An MPM describes a Markovian stochastic process $\vX(t)$ with $r$ competing Poissonian (memoryless) transitions $\vX\longrightarrow\vX+\vnu_j$, with rates $W_j(\vX)$. Its analytic formulation is a `master equation' for the probability mass $P(\vX;t)$:
\begin{equation} \label{e:ME}
  \partial_t P(\vX;t) = \sum_{i=1}^r \left\{ W_i(\vX-\vnu_i) P(\vX-\vnu_i;t) - W_i(\vX) P(\vX;t)\right\}.
\end{equation}

\subsection{A self-repressing gene network}\label{s:genex}
We introduce now a simple `running' example to illustrate the main concepts of the paper. Specifically, we consider the simplest gene network, composed of a single gene repressing its own expression. Despite its simplicity, this system is ubiquitously present in the genome \cite{Stelling2010}. We model it by a PCTMC $\calX=(\vX, {\mathcal M}, {\mathcal T}, \vX_0)$ with three variables, $\vX=(X_1,X_2, X_3)$, counting the amounts of, respectively, the repressed gene ($X_1$); the active, protein-producing gene ($X_2$); and the protein ($X_3$). The transcription-translation is lumped in one single step.  The state space is $\mathcal{M} = \{0,\ldots,N\}\times\{0,\ldots,N\}\times\NN$, where $N$ is the number of copies of the gene in the system (cf. also the discussion at the end of Section \ref{s:mf}).  The dynamics of the model is given by four transitions:
\begin{itemize}
\item $\tau_1 = (\makebox[1.25cm][l]{produce},\vnu_1 = (\phantom{+}0,0,\phantom{+}1)^\tsp,W_1(\vX) = \eps k_pX_2)$ -- protein production;
\item $\tau_2 = (\makebox[1.25cm][l]{degrade},\vnu_2 = (\phantom{+}0,0,-1)^\tsp,W_2(\vX) = \eps k_dX_3)$ -- protein degradation;
\item $\tau_3 = (\makebox[1.25cm][l]{repress},\vnu_3 = (-1,1,\phantom{+}0)^\tsp,W_3(\vX) = k_b X_2X_3/N)$ -- repression, caused by the protein binding to a gene;
\item $\tau_4 = (\makebox[1.25cm][l]{unbind},\vnu_4 = (1,-1,0)^\tsp,W_4(\vX) = k_uX_1)$ -- the unbinding event.
\end{itemize}
Two remarks are in order: first, we do not remove a protein from the system when it bounds to the repressor. This is a minor tweak that simplifies the following discussion. Secondly, as typical for bimolecular reactions \cite{Gillespie1977JPC}, we rescale the binding rate by the volume $N$, which for simplicity we assume here to equal the total amount of genes. In this way,  the (copy number) concentration  of the gene is between zero and one.

\section{The mean field limit of a MPM}\label{s:mf}

Consider a MPM for a fixed \emph{system size} $N$. The system size is usually interpreted as either the total population (typical of ecology and queueing networks' applications), or volume (chemical reaction networks). We can easily define a normalised MPM, by dividing variables by $N$, $\vX^N = \vX/N$, and expressing rates and updates with respect to these new variables. We call ${\mathcal{M}}^N$ the normalised state space, and further assume that the normalised state space satisfies $\bigcup_{N\in\NN} \mathcal{M}^N \subseteq E$ for some open set $E\subseteq \RR^n$. We call  $W_j^N:E\mapsto\RR_{\geq 0}$ the normalised rate functions for system size $N$, and assume $W_j^N(\vx)$ is defined for each $\vx\in E$ (as usually the case).
\begin{assumption} We  require that: \vspace{-1ex}
  \begin{enumerate}
  \item[(a)] For each $j=1,\ldots,r$, uniformly  for $\vx\in E$ it holds that 
    \begin{equation}\label{e:limitrate}
      w_j(\vx) = \lim_{N\to\infty} {W^N_j(\vx)}/{N}.
    \end{equation}
  \item[(b)] Smoothness of functions $w_j(\vx)$,   at least locally Lipschitz continuous.
  \item[(c)] The normalised initial conditions converge: $\vX^N_0\to\vx_0\in E$.
  \end{enumerate}
\end{assumption}
Under this assumption, the sequence of MPM $\vX^N(t)$ converges (in probability, for any finite time horizon) to the solution $\vx(t)=\vx(t,\vx_0)$ of the initial value problem
\begin{equation}\label{e:mf}
  \frac{\rmd \vx}{\rmd t}(t) =F(\vx(t))\,,\quad \vx(0)=\vx_0\,,\quad F(\vx)=\sum_{i=1}^r \vnu_i w_i(\vx),
\end{equation}
where $F(\vx)$ is  the (mean field) \emph{drift} of the MPM. More formally, the following theorem holds  \cite{EthierKurtz2005}:
\begin{theorem}
  \label{th:Kurtz}
  Under  conditions a, b, and c above, for any $T<\infty$ and $\eps >0$,
  \begin{equation*}
   \lim_{N\to\infty} \PP{ \sup_{t\le T} \norm{ \vX^N(t) - \vx(t) } > \eps}= 0.
  \end{equation*}
\end{theorem}
We stress that Theorem \ref{th:Kurtz} holds for any finite time window but it does not address the important question of steady state behaviour ($T=\infty$). Here the phenomenology is much wilder, and few things are known with certainty. However, if the mean field ODE (\ref{e:mf}) has a unique, globally attracting steady state $\vx(\infty)$, i.e. for each $\vx_0\in E$, $\lim_{t\to\infty}\vx(t,\vx_0) = \vx(\infty)$, then we have \cite{benaim_deterministic_2003,benaim_mean_2011,Bortolussi2013}:
\begin{theorem}
  \label{th:ssmf}
  Under the conditions of Theorem \ref{th:Kurtz}, if  $\vX^N(t)$ is ergodic and $\vx(t,\vx_0)$ has a unique globally attracting steady state, then
  \begin{equation*}
    \lim_{N\to\infty} \vX^N(\infty) = \delta_{\vx(\infty)}\ \ \text{in probability,}
  \end{equation*}
  where $\delta_{\vx(\infty)}$ is the point-wise mass probability at  $\vx(\infty)$.
\end{theorem}

\subsubsection{Running Example.}
The mean field equations for the simple gene model are 
\begin{equation*}
  \frac{\rmd x_2}{\rmd t}(t)=-\frac{\rmd x_1}{\rmd t}(t) = k_b x_2 x_3  - k_u x_1\,,\qquad\frac{\rmd x_3}{\rmd t}(t)=\eps k_p x_2 - \eps k_d x_3\,.
\end{equation*}
Theorem \ref{th:Kurtz} asserts that a solution of these ODEs is exactly equivalent to the corresponding MPM in the limit $N=\infty$. The important question is whether this ODE is an acceptable approximation when $N<\infty$, as is always the case in practice. 
Intuitively, if there are many (paralogue) copies of the gene, so that transcription can happen concurrently, this ODE may be expected to be an excellent approximation to the MPM with a finite, but large $N$. If the number of gene copies remains small and constant with respect to $N$, we can still construct a hybrid limit, see \cite{bortolussi_limit_2010}. For a discussion about the accuracy of mean field approximation, see \cite{Bortolussi2013}.

\section{Quasi-Equilibrium reduction}\label{s:qe}

In this section we provide formal definitions of the \emph{Quasi-Equilibrium} framework with two objectives in mind. Firstly, we aim at generalizing the `canonical' setting where the fast and slow components of a model are decoupled by premise. We assume that they could be entangled, paying the price of a little extra formality. The second goal is to present a formal guideline of reducibility in the form of a list of easily verifiable conditions. This is achieved in  Assumption \ref{a:qe_sep} of \sref{s:f-s}. However, we start by recalling two key ingredients of the reduction, \emph{coordinate transforms} and \emph{stoichiometric invariants} applied to MPMs.

\subsection{Image of a MPM under a change of coordinates}\label{s:ct}
A linear operator $L$ acting on a finite dimensional vector space ${\mathcal M}$ is equivalent to matrix multiplication. We would like describe the $L$-action on an MPM. Define $L_{\sf A}(\vx)={\sf A}^\tsp\cdot \vx$, where $\sf A$ is a real $n\times m$ matrix, and $\vx\in{\mathcal M}$. If, in addition, $\vy=L_{\sf A}(\vx)$ is invertible (${\sf A}$ is a square, invertible matrix) then the inverse, denoted by $\vx=L^{-1}_{\sf A}(\vy)$, is unique and $L^{-1}_{\sf A} = L_{{\sf A}^{-1}}$.

Fix such an invertible $L$ and consider an MPM $\calX=(\vX, {\mathcal M}, {\mathcal T}, \vX_0)$. The $L$-\emph{image} of $\calX$ is defined as $\calX_L:=L\circ\calX=(\vY, {\mathcal N}, {\mathcal T}, \vY_0)$, where
\begin{itemize}
\item $\vY = L(\vX)$, $\vY_0 = L(\vX_0)$, and ${\mathcal N} = L({\mathcal M})$;
\item Each transition $\tau = (\vnu,W)$ of $\calX$ becomes the transition $\tau' = (\vmu,W')$, where $\vmu = L(\vnu)$ and $W'(\vy) = W( L^{-1}(\vy))$.
\end{itemize}

It is obvious that, as $L$ preserves all the update rules $\vX\longrightarrow\vX+\vnu_j$, $\calX_L$ is equivalent to $\calX$, in a sense that $\calX_L$ represents the same stochastic process as $\calX$, viewed in transformed coordinates $\vY=L(\vX)$.

\subsection{Image of a MPM under  a stoichiometry reduction}\label{s:sr}
The $n\times r$ \emph{stoichiometry matrix} $\St$ of a MPM $\calX$ is a matrix composed from all the state change vectors $\vnu$, arranged as columns:
\begin{equation}\label{e:St}
  \St_\calX=\begin{pmatrix}\vnu_1, \dots, \vnu_r\end{pmatrix}\,.
\end{equation}
Two important characteristics of $\St$ are, the rank $\rank{(\St)}$, and the co-dimension 
\begin{equation}\label{e:codim}
\codim{(\St)}:=n-\rank{(\St)} \geq \max\{0,n-r\}\,.
\end{equation}

A MPM $\calX$ is called \emph{(stoichiometry) reducible} iff $m_\calX:=\codim{(\St_\calX)}>0$. By definition, there exist $m_\calX$ linearly independent vectors ${\bf c}_1$, \dots, ${\bf c}_{m_\calX}$ such that $L_{{\bf c}_i}(\vnu_j) = 0$ for all $i,j$. This implies, for each $Y_i=L_{{\bf c}_i}(\vX)$, a transition $Y_i \longrightarrow Y_i + L_{{\bf c}_i}(\vnu_j) = Y_i$. Therefore, the vector $\vY=(Y_1,\dots,Y_{m_\calX})$ is conserved by dynamics. Its components are called \emph{p-invariants}. They maintain constant values throughout dynamics therefore they can be made into parameters, rather than remaining independent variables. To achieve this, fix additional $n-m_\calX$ vectors $\vec{k}_1,\ldots,\vec{k}_{n-m_\calX}$, requiring that $\{\vec{c}_i\}$ and $\{\vec{k}_j\}$ should span ${\mathcal M}$. We arrange those vectors in two matrices ${\sf C} = \begin{pmatrix}\vec{c}_1, \dots, \vec{c}_{m_\calX}\end{pmatrix}$ and ${\sf K} = \begin{pmatrix}\vec{k}_1, \dots, \vec{k}_{n-m_\calX}\end{pmatrix}$. The matrix $\begin{pmatrix}{\sf C},{\sf K}\end{pmatrix}$ is then invertible by definition. The \emph{(stoichiometry) reduced} image of $\calX$ is defined as $\calX_{\sf C,K}:=(\vZ,{\mathcal K}, {\mathcal T}, \{ \vZ_0,\vY_0\})$ where
\begin{itemize}
\item $\vZ=L_{\sf K}(\vX)$,  $\mathcal{K} = L_{\sf K}(\mathcal{M})$, $\vY _0= L_{\sf C}(\vX_0)$, and $\vZ=L_{\sf K}(\vX_0)$;
\item Each transition $\tau = (\vnu,W)$ of $\calX$ becomes the transition $\tau = (\vsg,\overline{W}_{\vY_0})$, where $\vsg = L_{\sf K}(\vnu)$ and $\overline{W}_{\vY_0}(\vZ) = W\big( L^{-1}_{({\sf C,K})}(\vY_0,\vZ)\big)$.
\end{itemize}

\subsubsection{Running example.} Going back to the example of \sref{s:genex}, we have 
\[ \St = \big( \vnu_1,\vnu_2,\vnu_3,\vnu_4\big)=\begin{pmatrix*}[r]  0 & 0 & -1 & 1\\ 0 & 0 & 1 & -1\\ 1 & -1 & 0 & 0  \end{pmatrix*}. \]
We may recognise that $\vec{c} = (1,1,0)^\tsp$ is a p-invariant of the system. Letting $\vec{k}_1 = (0,1,0)^\tsp$, $\vec{k}_2 = (0,0,1)^\tsp$, we obtain the following reduced PCTMC model:
\begin{itemize}
\item $\vZ=(Z_1,Z_2)=(X_2,X_3)$, $Y_0= N$, $\mathcal{K} = \{0,\ldots,N\}\times \NN$;
\item State changes $\vsg_i$ are obtained from the corresponding $\vnu_i$s by crossing out the first element. Rates $\tilde{W}$ are equal to $W$s expressed in the new variables $\vZ$. The rate of the `unbind' transition becomes $\tilde W_4(Z_1,Z_2) = k_u (N-Z_1)$. 
\end{itemize}

\subsection{Fast-slow rate and variable decomposition of a PCTMC}\label{s:f-s}
We are now in position to describe the quasi-equilibrium reduction. 
\begin{assumption}\label{a:qe_sep}
Consider an MPM $\calX=(\vX, {\mathcal M}, {\mathcal T}, \vX_0)$ such that
\begin{enumerate}
\item[(a)] There exist two parameters $T^{\text{slow}}>T^{\text{fast}}>0$ and an integer $s$, $1<s<r$, such that the ordering of all rate functions
\begin{equation}\label{e:fs-rates}
  \underbrace{\displaystyle W_1(\vX)\le \dots \le W_s(\vX)}_{\hspace{-2mm}\displaystyle\text{slow transitions}} 
  \le\frac{N}{T^{\text{slow}}} < \frac{N}{T^{\text{fast}}} \le  
  \underbrace{W_{s+1}(\vX),\dots,W_r(\vX)}_{\hspace{2mm}\displaystyle\text{fast transitions}}
\end{equation}
is valid for all $\vX$ in a sufficiently large subspace of ${\mathcal M}$, containing the initial condition $\vX_0$. This condition is equivalent to requiring that rate functions behave with respect to dimensionless parameter $\eps = \frac{T^{\text{fast}}}{T^{\text{slow}}}$ as follows
\begin{eqnarray}
  W_i(\vX;\eps) &\underset{\eps\to0}{\sim}& \eps W_{0,i}(\vX) + O(\eps^2)\,,\quad  i=1,\dots, s \label{e:w-slow} \\
  W_i(\vX;\eps) &\underset{\eps\to0}{\sim}&      W_{0,i}(\vX) + O(\eps)\,,\quad i=s+1,\dots, r \label{e:w-fast}
\end{eqnarray}
where $W_{0,i}(\vX)$ are functions that do not depend on $\eps$.

The set of transitions ${\mathcal T}$ is thus partitioned into slow transitions ${\mathcal T}^{\text{slow}}=\{ \tau_1, \dots, \tau_s \}$, and fast transitions ${\mathcal T}^{\text{fast}}=\{\tau_{s+1},\dots,\tau_r\}$.

\item[(b)] $\calX$ restricted to ${\mathcal T}^{\text{fast}}$ is stoichiometry reducible according to \sref{s:sr}, i.e. 
  \begin{equation}
    m:=\codim{(\vnu_{s+1},\dots,\vnu_{r})} >0\,.
  \end{equation}
\end{enumerate}
\end{assumption}
If both these conditions are satisfied, we may separate slow and fast components of $\vX$, such separation being the basis of the subsequent dimensional reduction. Matrices ${\sf C}$ and $\sf K$ can be identified such that, following \sref{s:ct}, $({\sf C,K})$ is invertible and $L_{\sf C}(\vnu_i) = 0$, but only for $i=s+1,\dots,r$. Define 
\begin{equation}
  \underbrace{\vY = (Y_1,\dots, Y_m) = L_{\sf C}(\vX)}_{\displaystyle\text{slow variables}}   \quad 
  \underbrace{\vZ = (Z_1,\dots, Z_{n-m}) = L_{\sf K}(\vX)}_{\displaystyle\text{fast variables}}
\end{equation}
Note that $\vY$, owing to its definition in terms of fast transitions rather than all transitions, is not a p-invariant. This means that some transitions of the slow variable will occur, given by the updated vectors $\vmu$, defined as follows
\begin{equation*}
  L_{\sf C}\big( \vnu_1,\dots,\vnu_s,\vnu_{s+1},\dots,\vnu_r \big) = \big( \vmu_1, \dots, \vmu_s, {\bf 0}, \dots, {\bf 0}\big)\,.
\end{equation*}
The fast subspace update vectors are similarly defined: $\vsg_i=L_{\sf K}(\vnu_i)$.

\subsubsection{Running example.} 
We assume that $\eps\ll 1$ is a small dimensionless parameter. This assumption implies the partition ${\mathcal T}^{\text{slow}}=\{\tau_1,\tau_2\}$ and ${\mathcal T}^{\text{fast}}=\{\tau_3,\tau_4\}$. If all other parameters are $O(1)$, then there is a large gap between $T^{\text{fast}}$ and $T^{\text{slow}}$, guaranteed by the smallness of $\eps$, which we leave as the scale separation parameter.
The procedure of stoichiometry reduction, applied to ${\mathcal T}^{\text{fast}}$, provides 
\begin{equation*}
  m = \codim{\begin{pmatrix*}[r] 1 & -1 \\ 0 & 0 \end{pmatrix*}} = 2 - 1 = 1\,.
\end{equation*}
Since $m>0$, this model is QE-reducible and indeed, $\vec{c}=(0,1)^\tsp$ is a p-invariant of ${\mathcal T}^{\text{fast}}$. Complementing the basis with $\vec{k}=(1,0)$ we conclude that the slow variable, $Y=L_{\vec{c}}(\vX)=X_3$ is the protein, and the fast variable, $Z=L_{\vec{k}}(\vX)=X_2$ is the active gene. The $\eps$-rescaled rates $W_{0,i}$, expressed in the slow-fast variables, are 
\begin{equation} 
\label{e:runningRates}
W_{0,1} = k_p Z, \quad W_{0,2} = k_d Y, \quad W_{0,3}=k_b YZ/N, \quad W_{0,4}=k_u(N-Z).\end{equation}

\subsection{Quasi-Equilibrium reduction of the mean-field model}
As a demonstration of utility of our formalism, we will obtain the canonical equations of the singular perturbation theory \cite{Verhulst2005} from the standard quasi-equilibrium approximation of ODEs.

Recall from Section \ref{s:mf} the definition of limit rate functions (\ref{e:limitrate}) and  that of the limit  drift vector $\vec{F}(\vx)=\sum_{i=1}^r \vnu_i w_i(\vx;\eps)$, where we made explicit the dependence on a small parameter $\eps$. If the MPM satisfies Assumption \ref{a:qe_sep}, then the asymptotic $\eps\to0$ dependence of the rate functions is $w_i(\vx;\eps)\sim\eps w_{0,i}(\vx) + O(\eps^2)$ for $i=1,\dots, s$, and $w_i(\vx;\eps)\sim w_{0,i}(\vx) + O(\eps)$ for $j=s+1,\dots, r$ and $1<s<r$. Define the \emph{slow variables} $\vy=L_{\sf C}(\vx)$, the \emph{fast variables} $\vz=L_{\sf K}(\vx)$, and the \emph{slow time} $\tau=\eps t$. It is then straightforward to demonstrate that the mean field limit equations are equivalent to
\begin{equation}
\label{e:tik}
  \frac{\rmd\vy}{\rmd\tau} = G(\vy,\vz) + O(\eps),\quad  \eps\frac{\rmd\vz}{\rmd\tau} = H(\vy,\vz) + O(\eps) 
\end{equation}
where
\begin{equation}\label{e:GH}
  G=\sum_{i=1}^s L_{\sf C}(\vnu_i) w_{0,i}(L_{({\sf C,K})}^{-1}\big(\vy,\vz)\big)\,,\quad 
  H=\sum_{j=s+1}^r L_{\sf K}(\vnu_j) w_{0,j}(L_{({\sf C,K})}^{-1}\big(\vy,\vz)\big)\,.
\end{equation}
Since $\eps$ multiplies the highest order derivative in \eqref{e:tik} (right), the perturbation in $\eps$ is singular \cite{Verhulst2005}. The construction of a reduced model from equations \eqref{e:tik} is governed by further assumptions provided by the
\emph{Tikhonov theorem} \cite[Theorem 8.1]{Verhulst2005}. 
\begin{assumption}\label{a:tyc}
  Consider the initial value problem \eqref{e:tik} for $\tau\ge 0$, with $\vy(0)=\vy_0$, $\vz(0)=\vz_0$. We further require:
  \begin{enumerate}
  \item[(a)] the drifts $G(\vy,\vz)$ and $H(\vy,\vz)$ are  sufficiently smooth functions of their arguments.
  \item[(b)] a unique solution $\vy^\eps(\tau)$, $\vz^\eps(\tau)$ of the initial value problem \eqref{e:tik} exists;
  \item[(c)] a unique solution $\overline{\vy}(\tau)$, $\overline{\vz}(t)$ of the reduced initial value problem exists; the reduced problem being defined by
    \[ \rmd\vy/\rmd \tau = G(\vy,\vz), \quad \vy(0)=\vy_0,\quad 0 = H(\vy,\vz), \]
  \item[(d)] equation $0 = H(\vy,\vz)$ is solved by $\vz = {\bm\phi}(\vy)$ where $\bm\phi$ is continuous, and it is an isolated root; 
  \item[(e)] $\vz = {\bm\phi}(\vy)$ is an asymptotically stable solution of
    $\rmd\vz/\rmd t = H(\vy,\vz) $ uniformly in $\vy(\tau)$, considered as a (fixed) parameter;
  \item[(f)] $\vz(0)$ is contained in an interior subset of the domain of attraction of $\vz={\bm \phi}(\vy)$ for $\vy=\vy(0)$.	
  \end{enumerate}
\end{assumption}
\noindent The previous conditions guarantee that the solution of the reduced problem (defined in (c) above) is actually the $\eps\to 0$ limit of the original system, as proved in the following:
\begin{theorem}[Tikhonov (1958)] \label{t:tyc}
  Under conditions (a)--(f) above, $\forall T<\infty$
  \begin{equation}
    \lim_{\eps\to0}\vy^{\eps}(\tau) = \overline{\vy}(\tau), \quad\quad
        \lim_{\eps\to0}\vz^{\eps}(\tau) = \overline{\vz}(\tau),\quad 0 < \tau \le T	
  \end{equation}
\end{theorem}

\subsubsection{Running example.} In our example, the slow variable $y$ is the protein concentration, the fast variable $z$ is the active gene concentration. They satisfy \eqref{e:tik} in the slow time variable $\tau=\eps t$, with  $G(y,z)=  k_pz-k_d y$ and $H(y,z)= k_byz - k_u(1-z)$. Solving $H(y,z)= 0$ for $z$, we get $\phi(y) = \frac{k_u}{k_u+k_b y}$, hence finding the classic ODE for lumped gene transcription:
\[ \frac{d\bar{y}}{d\tau} =   \frac{k_pk_u}{k_u+k_b \bar{y}} - k_d \bar{y} \]

\subsection{The Quasi-equilibrium reduction of an MPM}
Let the Assumption \ref{a:qe_sep} hold for \eqref{e:ME} (the rate functions and the variable are decomposable into fast and slow subsets). Substiting the decomposition, described in \sref{s:f-s}, into \eqref{e:ME}, yields
\begin{eqnarray}\label{e:me}
  \partial_t P(\vY,\vZ;t) &=& \sum_{i=1}^r\Big\{ W_i(\vY-\vmu_i,\vZ-\vsg_i) P(\vY-\vmu_i,\vZ-\vsg_i;t) \nonumber \\
  && - W_i(\vY,\vZ)P(\vY,\vZ;t)\Big\}, \quad P(\vY,\vZ;0)=P_0(\vY,\vZ).
\end{eqnarray}
In addition to requiring that a corresponding MPM satisfies Assumption \ref{a:qe_sep}, we further require
\begin{assumption}{(Ergodicity)} \label{a:ergodicity}
  \begin{enumerate}
  \item[(a)] The full process $\vX(t) = (\vY,\vZ)(t)$ is ergodic; 
  \item[(b)] The stochastic process $\vZ_{\vY}(t)$ describing the fast subsystem  is ergodic for each fixed $\vY$.
  \end{enumerate}
\end{assumption}

\noindent Under these further requirements, the master equation of the reduced system is 
  \begin{equation}\label{e:me-slow}
    \partial_\tau P(\vY;\tau)=\sum_{i=1}^s\Big\{ \widetilde{W}^\infty_{0,i}(\vY-\vmu_i) P(\vY-\vmu_i;\tau) - \widetilde{W}^\infty_{0,i}(\vY) P(\vY;\tau)\Big\} 
  \end{equation}
  \begin{equation}\label{e:wavg}
    \widetilde{W}_{0,i}^\infty =  \EE_{\vZ_{\vY}(\infty)}( W_{0,i}(\vY,\vZ)) =  \sum_{\vZ} W_{0,i}(\vY,\vZ) \overline{P}_\vY(\vZ), \quad i=1,\dots,s
  \end{equation}
and $\vZ_{\vY}(\infty)$ is the unique steady state measure of the fast process $\vZ_{\vY}(t)$ (due to \ref{a:ergodicity}.b), with  $\overline{P}_{\vY}(\vZ)$ being the steady state probability of the master equation
\begin{equation}\label{e:me-fast}
  \partial_t P_\vY(\vZ;t)=\sum_{j=s+1}^r\Big\{ W_{0,j}(\vY,\vZ-\vsg_j) P_\vY(\vZ-\vsg_j;t) - W_{0,j}(\vY,\vZ) P_\vY(\vZ;t)\Big\}
\end{equation}
The slow process $\widetilde{\vY}(\tau)$ defined by the master equation \eqref{e:me-slow}  is indeed the limit of the full process for $\eps\to 0$ (see the appendix for the proof):
\begin{theorem}[Quasi-equilibrium reduction]\label{t:qe}
  Under assumptions \ref{a:ergodicity}.(a)--(b), 
  \begin{equation}
    \lim_{\eps\to0} \sum_{\vZ} P(\vY,\vZ;\tau/\eps) = P(\vY;\tau)
\end{equation}
for all $T>0$ and $0\le \tau \le T$. \qed
\end{theorem}
We can now lift Theorem \ref{t:qe} to the MPM level. Consider an MPM $\calX=(\vX, {\mathcal M}, {\mathcal T}, \vX_0)$, that is QE reducible (see \sref{s:f-s}). The quasi-equilibrium image of $\calX$ is defined as  $\calX^{\text{qe}}=(\vY,{\mathcal N}, {\mathcal T}^{\text{slow}}, \{\vY_0,\vZ_0\})$, where
\begin{itemize}
\item $\vY=L_{\sf C}(\vX)$, $\mathcal{N} = L_{\sf C}(\mathcal{M})$, $\vY _0= L_{\sf C}(\vX_0)$, and $\vZ_0=L_{\sf K}(\vX_0)$;
\item Each \emph{slow} transition $\tau = (a,\vnu,W)\in{\mathcal T}^{\text{slow}}$ of $\calX$ becomes the transition $\tau = (a, \vmu,\widetilde{W}_0^\infty)$, where $\vmu = L_{\sf C}(\vnu)$ and $\widetilde{W}_0^\infty(\vY)$ is defined by \eqref{e:wavg}.
\end{itemize}
It is also straightforward to define a \emph{family} of MPMs for the fast subsystem, parametrized by the slow variable $\vY$,  described by the master equation \eqref{e:me-fast}.

\subsubsection{Running example.}
The most important new information are the expressions for the averaged slow rates, given by the definition \eqref{e:wavg}. We find
\begin{equation}\label{e:genex-wavg}
  \widetilde{W}_{0,1}(Y) = \sum_Z k_p Z \overline{P}_Y(Z) = k_p \langle Z \rangle_{Z_Y(\infty)} \,,\quad 
  \widetilde{W}_{0,2}(Y) = k_p Y.
\end{equation}
These rates, together with the state change vectors $\mu_1=1$, $\mu_2=-1$, complete the definition of a reduced MPM, which is easaly seen to describe a birth-death process. The rates of this process are given by \eqref{e:genex-wavg}. In this simple case, the fast process, conditional on $Y$, is also a birth-death process, hence owing to the linearity of (\refeq{e:runningRates}), we get $\langle Z \rangle_{Z_Y(\infty)} = \frac{Nk_u}{k_u+k_bY/N}$, which gives an explicit expression for the rates of $Y$. 
We emphasize that in general this is not true, as the stationary distribution of $Z_Y$ may not be known explicitly, so that one has still to rely on numerical methods, like simulation \cite{weinan_nested_2007}.

\begin{figure}
  \includegraphics[width=0.3\linewidth]{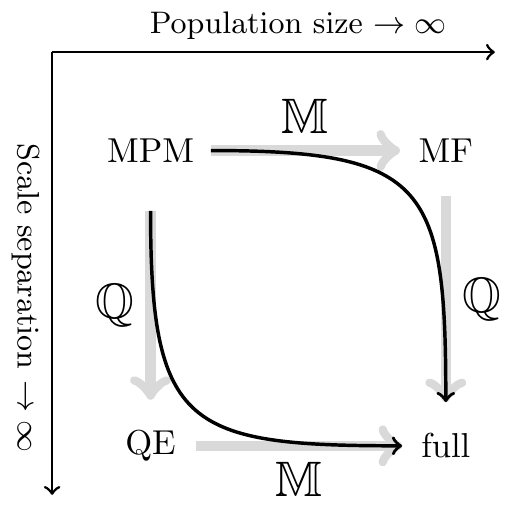}
  \includegraphics[width=0.68\linewidth]{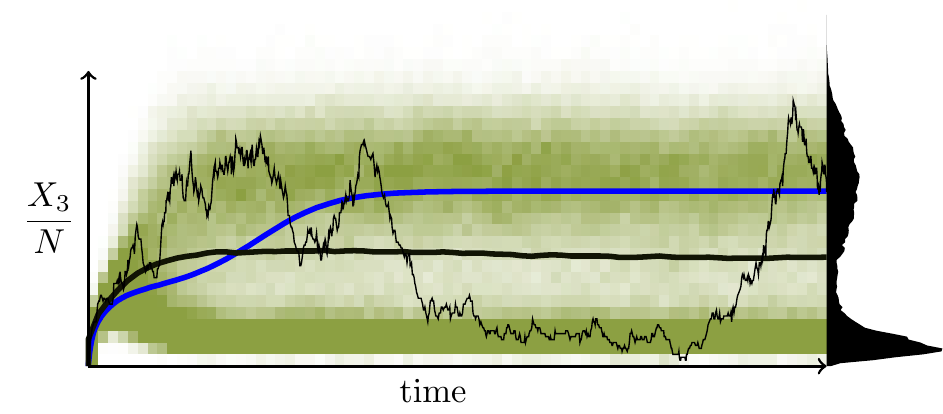}
  \caption{\label{f:diagram} Left: commutation diagram. The curved paths illustrate two distinct limiting procedures to arrive from an MPM to a fully reduced model.
Right: the toggle switch counter example to Theorem \ref{th:mfqe}. The blue curve is the solution of the reduced mean field ODE, while the solid black curve is the average of the reduced stochastic process, which is bistable (cf. the empirical distribution on the right). }
\end{figure}

\section{Comparing mean field and quasi-equilibrium}\label{s:comp}
Consider an MPM $\calX$, and, as in Section \ref{s:mf}, let $\vX^N(t)$ be the normalised model with respect to system size $N$. In this paper we have introduced  two possible model simplification strategies: the mean field approximation and the QE-reduction. To fix notation in the rest of this section, we will refer to the former by the operator $\mathbb{M}$, and to the latter by the operator $\mathbb{Q}$. Hence, $\mathbb{M}(\vX^N(t)) = \vx(t)$ is the mean field limit of $\vX^N(t)$, and $\mathbb{Q}(\vX^N(t)) = \widetilde{\vY}^N(\tau)$ is the QE reduction of $\vX^N(t)$, whereas $\mathbb{Q}(\vx(t)) = \widetilde{\vy}(\tau) = \mathbb{Q}(\mathbb{M}(\vX^N(t)))$ the QE reduction of $\vx(t)$.

The issue we wish to address in this section is how these two procedures are related. In particular, it is natural to ask if the two operators $\mathbb{M}$ and $\mathbb{Q}$ commute, as shown in Fig. \ref{f:diagram}. The diagram illustrates the following two possibilities. We could either construct the QE reduction upon the mean field limit of a MPM $\vX^N(t)$ and obtain a deterministic process $\widetilde{\vy}(\tau)$, or we could first apply the QE reduction to $\vX^N(t)$, and then attempt to construct the mean field limit of $\widetilde{\vY}^N(t)$. Two questions arise naturally
\begin{enumerate}
\item Does $\mathbb{M}(\mathbb{Q}(\vX^N(t)))$, i.e. the mean field limit of $ \widetilde{\vY}^N(\tau)$ exist? 
\item If so, is it the same as $\widetilde{\vy}(\tau) = \mathbb{Q}(\mathbb{M}(\vX^N(t)))$, i.e. does the diagram in Figure \ref{f:diagram} commutes?
\end{enumerate}
We show that the answer is `yes to both questions' only if some additional requirements for the fast substystem are fulfilled. We will demonstrate that the answer to question 2 is `no in general', and that even question 1 may have a negative answer. The problem is intimately connected with the extension of  Theorem \ref{th:Kurtz} to the steady state, hence with Theorem \ref{th:ssmf}. In fact, when we construct the QE reduction $\mathbb{Q}(\vX^N(t))$ of  $\vX^N(t)$, we need to average the slow rates with respect to the steady state distribution $\vZ_{\vY}^N(\infty)$ of the fast subsystem $\vZ_{\vY}^N$. Assumptions \ref{a:ergodicity}.(a) and \ref{a:ergodicity}.(b) enforce ergodicity, hence existence and uniqueness of such a steady state distribution $\vZ_{\vY}^N(\infty)$  for each $N$ and $\vY$. However,  to construct the mean field limit of  $\mathbb{Q}(\vX^N(t))$, we also need to know how   such sequence behaves as $N$ goes to infinity. Essentially, we need to know if it has a limit, and what such limit is. Unfortunately, this is one of the most delicate points of mean-field approximation theory: Little is known about the limiting behaviour of the steady state, except from Theorem \ref{th:ssmf}. Hence, we can provide a positive answer to questions 1 and 2 only if we place ourselves in the conditions of such a theorem. This leads to the following
\begin{assumption} 
\label{a:uniqueness}
The solution $\vz={\bm\phi}(\vy)$  of $0=H(\vy,\vz)$  is unique, i.e. the mean field limit of the fast subsystem $\bar{\vz}(t) = \bar{\vz}(t,\vy)$ has a unique, globally attracting equilibrium ${\bm \phi}(\vy)$  for  each value of the slow variables $\vy$. 
\end{assumption}
Under this assumption, we can apply Theorem \ref{th:ssmf} and conclude that, for each $\vY$, it holds that
$\vZ_{\vY}^N(\infty) \rightarrow \delta_{{\bm\phi}(\vY)}$ in probability.
At this stage, however, we need a further technical assumption (see also Remark \ref{rem:assumption}):
\begin{assumption}
  \label{a:uniformConvergenceSS}
  $\vZ_{\vY}^N(\infty)$ converges to $\delta_{{\bm\phi}(\vY)}$ uniformly in $\vY$, i.e. $\forall \eps>0$, 
  \[\lim_{N\rightarrow \infty}\PP{\sup_{\vY\in \mathcal{N}^N} \|\vZ_{\vY}^N(\infty) -{\bm\phi}(\vY)\| > \eps } = 0.\]
\end{assumption}
Under these two additional assumptions, it is easy to show that 
\begin{equation}
\frac{\widetilde{W}_{0,i}^{\infty}(\vy)}{N} \underset{N\to\infty}{\longrightarrow} w_{0,i}(\vy)
\end{equation}
uniformly in $\vy$.  This readily implies that the drift of the QE-reduced process $\widetilde{\vY}^N(t)$,  $\widetilde{F}^N(\vy):=\sum \vnu_i \frac{\widetilde{W}_{0,i}^{\infty}(\vy)}{N}$ converges uniformly to the drift $G(\vy,{\bm\phi}(\vy))$, defining the vector field of the QE-reduced mean field limit, as in equation (\ref{e:tik}), which is sufficiently regular by hypothesis \ref{a:tyc}.(a). Hence, the conditions of Theorem \ref{th:Kurtz} are satisfied by the sequence of processes $\widetilde{\vY}^N(t)$, and we can conclude that 
\begin{theorem}
  \label{th:mfqe}
  Under Assumptions \ref{a:uniqueness} and \ref{a:uniformConvergenceSS} above, with $T<\infty$ fixed and for each $t\leq T$, $\mathbb{M}(\mathbb{Q}(\vX^N(t)))$ exists and $\mathbb{M}(\mathbb{Q}(\vX^N(t)))=\mathbb{Q}(\mathbb{M}(\vX^N(t)))$ . \qed
\end{theorem}
\begin{remark}\label{rem:assumption}
  Assumption 2 requires that the convergence of the sequence of steady state measures  of the  fast subsystem to their limit point-wise distribution is uniform in the slow state $\vY$.  We conjecture this is in fact true without any further requirement on the MPM. A heuristic argument goes as follows: by the functional central limit \cite{EthierKurtz2005}, we know that the fast subsystem will behave like a Gaussian process for $N$ large enough. In particular, the steady state distribution of  $\vZ_{\vY}^N(\infty)$  will be approximatively Gaussian with mean ${\bm\phi}(\vY)$ and Covariance matrix $ C^N(\vY) = \frac{1}{\sqrt{N}}C(\vY)$, where $C(\vY)$ does not depend on $N$ and it is the steady state solution of the covariance linear noise equations \cite{Gardiner1985}. As such, it will depend continuously on $\vY$. Using similar arguments as in the proofs of Kurtz theorem, we can guarantee that the eigenvalues of $C(\vY)$ are uniformly bounded by a constant $\Lambda<\infty$, which implies that we can find a uniform bound in $\vY$ on the spread of the steady state distribution, going to zero as the population size $N$ diverges. A formal proof of Assumption \ref{a:uniformConvergenceSS} seems to be strictly related to the availability of explicit bounds for the convergence in probability of $\vZ_{\vY}^N(\infty)$ to $\delta_{{\bm\phi}(\vY)}$, which is still an open issue, see also \cite{BH13}.
 \end{remark}
 
 \subsubsection{Running example.} The mean field equation for the fast variable $z$ in the self-repressing gene example is linear, so that  it is easy to see that it has a unique globally attracting equilibrium for each $y$. Furthermore, for any $N$ and $y$, it holds that ${\widetilde{W}_{0,1}^{\infty}(y)}/{N} = w_{0,1}(y)$ (cf. the expression of ${\widetilde{W}_{0,1}^{\infty}(y)}$ computed at the end of last section), hence Assumption \ref{a:uniformConvergenceSS} is trivial in this case. Therefore Theorem \ref{th:mfqe} applies: mean field and time scale reduction commute.

\subsection{On the necessity of Assumption \ref{a:uniqueness}}
Assumption \ref{a:uniqueness}, on the other hand, is quite crucial for Theorem \ref{th:mfqe} to hold. Without it, we cannot say much about the limit behaviour of the sequence of steady state measures of the fast subsystem, a part from the fact that each limit point will be supported in the Birkhoff center of the limit mean field dynamical system \cite{Bortolussi2013,benaim_deterministic_2003}. If this system has only stable and unstable equilibria as invariant sets (e.g. it satisfies the conditions of \cite{boudec_stationary_2013}), then each limit point of the sequence of steady state measure will be supported in those equilibria, but this is as much as we can say. In particular, \emph{we cannot guarantee the existence of a  limit} for such a  sequence, hence \emph{the reduced stochastic model may not be amenable of mean field approximation}.  However, we can argue that, in case the limit of $\vZ_{\vY}^N(\infty)$ is defined, then $\mathbb{M}$ and $\mathbb{Q}$ will not generally commute. The reason for this is to be found in the large deviations theory for (population) CTMC \cite[Ch.~6]{shwartz_large_1995}, which guarantees that each trajectory of the stochastic system will remain close to all stable equilibrium of the mean field limit a non-negligible fraction of time. Hence, the limit steady state measure, if any, must be a \emph{mixture of pointwise masses} concentrated on (stable) equilibria\footnote{The role of unstable equilibria is unclear. It is plausible that they will be visited only for a vanishing fraction of time, but we know no proof of this fact.}.  On the other hand, the fast subsystem $\bar{\vz}$ of  the mean-field limit will converge to a single stable equilibrium  (assuming no bifurcation event happens in the  fast subsystem as $\bar{\vy}(t)$ varies, i.e. that Assumption \ref{a:tyc}.(e) is in force). This implies that the limit for $N\rightarrow\infty$ of the rates $\frac{\widetilde{W}_{0,i}^{\infty}(\vy)}{N}$ will not converge to $w_{0,i}(\vy)$, which is evaluated on  the single equilibrium $\vz={\bm\phi}(\vy)$, but rather to a weighted average of the rate function $w_i$ evaluated on all (stable) equilibria.

To render this discussion more concrete, we illustrate this phenomenon by means of a genetic network model of a toggle switch \cite{Gardner2000}. We have three protein species, whose number is given by variables $\vX=(X_1,X_2,X_3)$, living in a volume $N$, with density $x_j = X_j/N$ (possibly exceeding unit value). The MPM is specified by the following six transitions: 
\begin{description}
\item \makebox[3.1cm][l]{Production of $X_1$~:} $\vnu_1 = (\phantom{+}1,0,0)^\tsp$,\,\,$W_1(\vX) =  \alpha_1 N^{\beta_1+1}/\big(N^{\beta_1} + X_2^{\beta_1}\big)$;
\item \makebox[3.1cm][l]{Degradation of $X_1$~:} $\vnu_2 = (-1,0,0)^\tsp$,\,\,$W_2(\vX) =  X_1$;
\item \makebox[3.1cm][l]{Production of $X_2$~:} $\vnu_3 = (0,\phantom{+}1,0)^\tsp$,\,\,$W_3(\vX) =  \alpha_2 N^{\beta_2+1}/\big(N^{\beta_2} + X_1^{\beta_2}\big)$;
\item \makebox[3.1cm][l]{Degradation of $X_2$~:} $\vnu_4 = (0,-1,0)^\tsp$,\,\,$W_4(\vX) =  X_2$;
\item \makebox[3.1cm][l]{Production of $X_3$~:} $\vnu_5 = (0,0,\phantom{+}1)^\tsp$,\,\,$W_5(\vX) =  \eps X_1$;
\item \makebox[3.1cm][l]{Degradation of $X_3$~:} $\vnu_6 = (0,0,-1)^\tsp$,\,\,$W_6(\vX) =  \eps X_3$;
\end{description}
The proteins `1' and `2' mutually repress each other, and thus properly constitute the toggle switch. Molecule `3' instead, is a slow product of the protein `1', and does not influence the toggle switch. This example is cooked up so that if $\eps\ll1$ then the variable $x_3$ and transitions $\tau_5$ and $\tau_6$ are trivially the slow ones. It should still be possible to see the breakdown of the assumptions 5 \& 6 in the long time expectation value of the molecule `3'. First we consider the mean field limit   
\[
  \frac{\rmd x_1}{\rmd t} = \frac{\alpha_1}{1+ x_2^{\beta_1}} - x_1\,,\qquad  \frac{\rmd x_2}{\rmd t} = \frac{\alpha_2}{1+ x_1^{\beta_2}} - x_2\,,\qquad \frac{\rmd x_3}{\rmd t} = \eps (x_1 - x_3).
\]
For a symmetric toggle model with parameters $\alpha_1=\alpha_2 =10$, $\beta_1=\beta_2 =1.4$, the two stable equilibria are $(\overline{x}_1,\overline{x}_2)=(a,b)$, $(b,a)$ where $a=0.764$, $b=5.931$. The limiting behavior of $x_3$ is $x_3 \underset{\tau\to\infty}{\longrightarrow} \overline{x}_1$, where $\overline{x}_1$ is either $a$ or $b$, depending on whose basin of attraction covers the initial condition of the trajectory. We took the initial conditions that are below the diagonal $x_1=x_2$. Such initial conditions are attracted to the equilibrium $\overline{x}_1=b$. The mean field time series $x_3$ vs $t$ is displayed in figure \ref{f:diagram}, where the mean field trajectory saturates at $b$ (blue curve).

Next we consider the stochastic dynamics. A representative stochastic time series of $X_3/N$ vs $t$ is shown in figure \ref{f:diagram}. Its variations are wider than a Gaussian approximation of the probability would imply. Sufficient insights can be gained by looking at the expectations of the form $\avg{\vX}(t) = \sum_{\vX} \vX P(\vX;t)$. The expectation of molecule `3' satisfies an exact differential equation 
$$ \rmd \avg{X_3}/\rmd t = \eps\avg{X_1} - \eps\avg{X_3}$$
Making a QE approximation to this equation is equivalent to replacing $\avg{X_1}$ with the equilibrium expectation $\overline{X}_1^\infty$ of the fast (`1+2') subsystem, and $\avg{X_3}(t)$ -- with $\widetilde{X}_3(\tau)$, each of which should be expressed in terms of their respective reduced probabilities. Since $X_3$ is decoupled from $X_1$ in the full model, $\overline{X}_1(t)=\avg{X_1}(t)$. Moreover, if $\eps\ll1$, we can also take $\overline{X}_1^\infty\approx\avg{X_1}(t)$, resulting in 
$$ \rmd \widetilde{X}_3(\tau) /\rmd\tau = \overline{X}^\infty_1 - \widetilde{X}_3(\tau). $$
Within this approximation, the solution tends to $\widetilde{X}_3(\tau) \underset{\tau\to\infty}{\longrightarrow} \overline{X}_1^\infty$. Then, comparison of ${x}_3(t)$, obtained from the mean field limit, and $\widetilde{x}_3(\tau)=\widetilde{X}_3(\tau)/N$, obtained from the stochastic model, provides a good measure of differences between the two approximations. The mean field trajectory, discussed in the previous paragraph, should be compared with the expectation $\widetilde{X}_3(\tau)$, shown as a solid gray line in figure \ref{f:diagram}. There is a significant difference between the two, suggesting the non-equivalence of reduced models in this particular case. Applying large deviations arguments \cite[Ch.~6]{shwartz_large_1995}, one may  expect $P(X_3)$  (shown as a density in figure \ref{f:diagram}) to look like, as $N\to\infty$, a mixture of point masses, concentrated  equilibria. Conjecturing that the mass is distributed only on stable equilibria and owing to the symmetry between $X_1$ and $X_2$, such weights will be equal to $\frac{1}{2}$, so $\widetilde{x}_3(\infty)=(a+b)/2$. A simulation supports this conjecture, as the curve for $\widetilde{X}_3(\tau)$ is roughly in the middle between the two peaks of the probability density shown in \ref{f:diagram}. 

\section{Discussion}\label{s:dis}

In this paper, we discussed in a homogeneous way two  approximation techniques for Markov Population Models: the mean-field limit and the quasi-equilibrium reduction  in the presence of multiple time-scales. Both approaches are based on  a notion of limit: for large population in the former case, and for a diverging separation of time scales in the latter. Our first contribution of this paper is to formalise in a clear way the quasi-equilibrium reduction for MPM, proving also the convergence of the original model to the  reduced one in the stochastic setting. The second original ingredient of this work is the investigation of the relationship between QE and mean-field. In particular, we identified sufficient conditions under which the two limits commute. We also argued that the commutation should not hold in general. The situation here is intimately connected with the nature of mean field convergence for steady state distributions. 

The take-home message is that care must be exercised when time scale separation techniques are combined with mean field limits. The behaviour of the system that we obtain by first taking the mean field limit and then the QE-reduction,  the most common way in literature, may not reflect at all the actual behaviour of the original stochastic model. Hence, one has to additionally show that the fast subsystem is well behaved (i.e., it satisfies assumption  \ref{a:uniqueness}). 

We note here that most of the assumptions we introduced hold in almost all practical cases, and are generally easy to verify. The most challenging ones are the separation of time scales (Assumption \ref{a:qe_sep}), and those related to the steady state behaviour of ODE models, i.e. Assumptions \ref{a:uniqueness} and \ref{a:tyc}.(e).

This line of research can be extended in few directions. First of all, the literature on time scale separation for  MPM is not as well developed as the literature for ODE models \cite{BortPaskSurvey14}. Many ideas developed in this  context can possibly be exported to MPM, especially techniques that automatically identify multiple time scales \cite{lam}. Finally, we are investigating how QE reduction propagates to moment closure-based approximations of variance and of higher order moments of the stochastic population  process.


\clearpage
\appendix
\section{Proofs}\label{s:app1}


\subsection{Proof of Theorem \ref{t:qe}}\label{s:thm1}
For simplicity, we prove the theorem under the assumption that the state space the full MPM is finite.
 
The slow variable probability, defined by marginalising the joint probability, $P^\eps(\vY;\tau)=\sum_{\vZ} P^\eps(\vY,\vZ;\tau)$, satisfies an equation, obtained by marginalising the master equation. Using the identity $\sum_{\vZ}W_iP^\eps(\vX,\vY)=P^\eps(\vY)\sum_{\vZ}W_iP^\eps_\vY$ for the rates from the slow transition subset we estimate
$(1/\eps)W_i(\vY,\vZ)P^\eps_\vY(\vZ;t)=W_{0,i}(\vY,\vZ)\overline{P}_\vY^\infty(\vZ) + W_{0,i}(\vY,\vZ) (P^\eps_\vY(\vZ;t) - \overline{P}_\vY^\infty(\vZ)) + O(\eps)$, while the fast rate subset terms are cancel out by the equal positive and negative contributions to the master equation. Finally, we get
\begin{eqnarray}
  \partial_\tau P(\vY;\tau) &=& \sum_\vZ \eps^{-1}\partial_t P(\vY,\vZ;\tau/\eps) \nonumber \\
  & = & \sum_{i=1}^s\Big\{ \widetilde{W}^\infty_i(\vY-\vmu_i) P^\eps(\vY-\vmu_i;\tau) - \widetilde{W}_i^\infty(\vY) P^\eps(\vY;\tau) \label{e:slowVarsFull} \\
  & & + \widetilde{D}^\eps_i(\vY-\vmu_i;\tau) - \widetilde{D}^\eps_i(\vY;\tau) \Big\} \nonumber
\end{eqnarray}
where the rates $\widetilde{W}^\infty_i$ are defined  by \eqref{e:wavg}, and 
\begin{equation}
  \widetilde{D}^\eps_i(\vY;\tau)=\sum_{\vZ}W_{0,i}(\vY,\vZ) \big(P^\eps_{\vY}(\vZ;\tau/\eps) - \overline{P}_{\vY}^\infty(\vZ)\big) + O(\eps)
\end{equation}

For the fast variable conditional probability $P^\eps_\vY(\vZ;t)$, differenting the identity $P(\vY,\vZ;t) = P(\vY;t)P_\vY(\vZ;t)$, we get
\begin{equation}\label{e:me-fast1}
  P(\vY;t) \partial_t P_\vY(\vZ;t) = \partial_t P(\vY,\vZ;t) - P_\vY(\vZ;t) \sum_{\vZ'} \partial_t P(\vY,\vZ';t)
\end{equation}
which we wish to divide by $P(\vY;t)$. A possible obstacle of $P(\vY;t)=0$ occurring for some $\vY$ is excluded except, possibly, at $t=0$, by the ergodicity assumption \cite{Schnakenberg1976}. Thus, dividing by $P(\vY;t)$ is warranted, and we find
\begin{eqnarray}
  \partial_t P_\vY(\vZ;t) &=& \sum_{j=s+1}^r\Big\{ W_j(\vY,\vZ-\vsg_j) P_\vY(\vZ-\vsg_j;t) -W_j(\vY,\vZ) P_\vY(\vZ;t)\Big\} \nonumber \\
  && +\sum_{i=1}^s \eps\Big\{ \delta(\vY;\vmu_i)\Gamma_i(\vY-\vmu_i,\vZ-\vsg_i;\vY,\vZ;t) \nonumber \\
  && -\Gamma_i(\vY,\vZ;\vY,\vZ;t)\Big\}
\end{eqnarray}
where $\Gamma$ and $\delta$ are defined by
\begin{eqnarray}
  \Gamma_i(\vY',\vZ';\vY,\vZ;t) &=& P_{\vY'}(\vZ';t)W_{0,i}(\vY',\vZ') \nonumber \\
  && - P_\vY(\vZ;t)\sum_{\vZ''}W_{0,i}(\vY',\vZ'') P_{\vY'}(\vZ'';t) + O(\eps) \\
  \delta(\vY,\vmu_i;t) &=& \frac{P(\vY-\vmu_i;t)}{P(\vY;t)}\qquad i=1,\dots,s
\end{eqnarray} 
By ergodicity assumption of the complete and reduced systems, both $P^\eps_\vY(\vZ;t)$ and $P^0_\vY(\vZ;t)$ admit a unique steady state. Then by implicit function theorem, $\overline{P}^\infty_\vY(\vZ) - P^\eps(\vY;\tau/\eps) = O(\eps)$. 

Then, the perturbation $\widetilde{D}_i$ in \eqref{e:slowVarsFull} is also of the order $\eps$. This term converges  to zero (uniformly in $\vY$, due to the finiteness of the state space), hence the vector field defining the reduced master equation for $\vY$ converges to the one of equation  \eqref{e:me-slow}, which by smoothness of the functions involved,  proves the theorem.

\end{document}